\documentclass[letterpaper]{article} 
\usepackage{aaai2026}  
\usepackage{times}  
\usepackage{helvet}  
\usepackage{courier}  
\usepackage[hyphens]{url}  
\usepackage{graphicx} 
\usepackage{natbib}  
\usepackage{caption} 
\usepackage{algorithm}
\usepackage{algorithmic}
\usepackage{amsmath}
\usepackage{amsfonts}
\usepackage{multirow}
\usepackage{newfloat}
\usepackage{listings}
\DeclareCaptionStyle{ruled}{labelfont=normalfont,labelsep=colon,strut=off} 
\floatstyle{ruled}
\newfloat{listing}{tb}{lst}{}
\floatname{listing}{Listing}
\title{Modeling Trend Dynamics with Variational Neural ODEs for \\ Information Popularity Prediction}
\author{
    Yuchen Wang,
    Dongpeng Hou,
    Weikai Jing,
    Chao Gao\thanks{Corresponding author},
    Xianghua Li,
    Yang Liu
}
\affiliations{
    Northwestern Polytechnical University\\
    \{wany810, hdp0918, 2015217442\}@mail.nwpu.edu.cn,
    \{cgao, li\_xianghua\}@nwpu.edu.cn, yangliuyh@gmail.com
}

\usepackage{bibentry}

\begin{document}

\maketitle

\begin{abstract}
Predicting the future popularity of information in online social networks is a crucial yet challenging task, due to the complex spatiotemporal dynamics underlying information diffusion. Existing methods typically use structural or sequential patterns within the observation window as direct inputs for subsequent popularity prediction. However, most approaches lack the ability to explicitly model the overall trend of popularity up to the prediction time, which leads to limited predictive capability. To address these limitations, we propose VNOIP, a novel method based on variational neural Ordinary Differential Equations (ODEs) for information popularity prediction. Specifically, VNOIP introduces bidirectional jump ODEs with attention mechanisms to capture long-range dependencies and bidirectional context within cascade sequences. Furthermore, by jointly considering both cascade patterns and overall trend temporal patterns, VNOIP explicitly models the continuous-time dynamics of popularity trend trajectories with variational neural ODEs. Additionally, a knowledge distillation loss is employed to align the evolution of prior and posterior latent variables. Extensive experiments on real-world datasets demonstrate that VNOIP is highly competitive in both prediction accuracy and efficiency compared to state-of-the-art baselines.
\end{abstract}

\begin{links}
    \link{Code}{https://github.com/cgao-comp/VNOIP}
\end{links}

\section{Introduction}
\begin{figure}[ht]
    \centering
    \includegraphics[width=1.\linewidth]{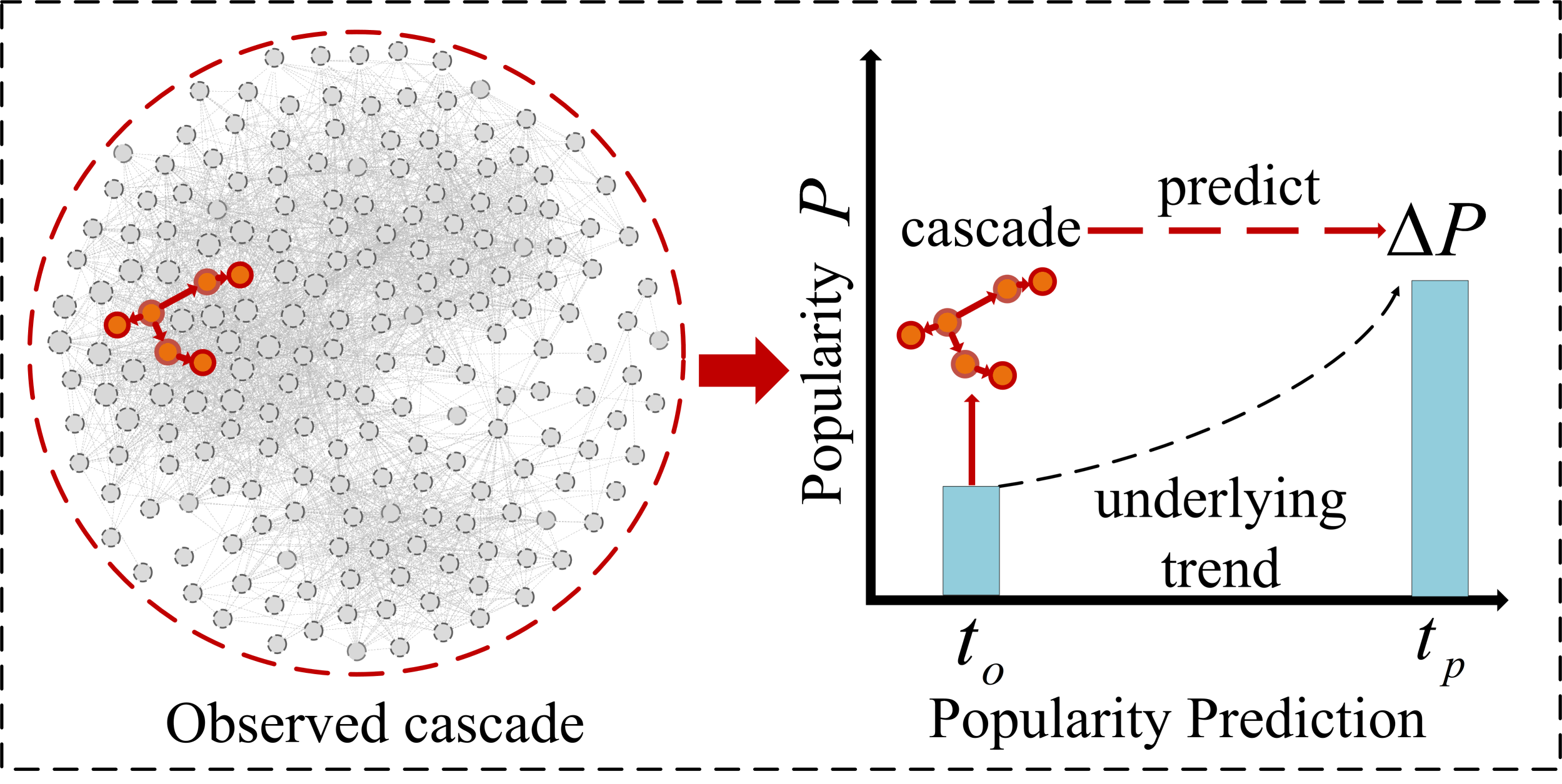}
    \caption{An illustration of popularity prediction.}
    \label{fig:1}
\end{figure}

With the explosive growth of social media platforms, understanding and forecasting the popularity of information has become an essential problem in social network services (SNS)~\cite{li2021capturing}. Information cascade popularity prediction focuses on modeling the overall diffusion scale and growth trajectory of information, rather than merely tracking individual behaviors, as illustrated in Fig.~(\ref{fig:1}). This macroscopic perspective is critical for grasping the collective dynamics and emergent patterns in large-scale social systems.
Moreover, accurately estimating the future reach and influence of content is vital for a wide range of real-world SNS applications, including viral marketing~\cite{rizoiu2017online}, event detection~\cite{steuber2023real}, and content recommendation~\cite{wang2018learning}.
Therefore, cascade popularity prediction stands as a fundamental and challenging problem and has attracted increasing attention.

Mainstream approaches to popularity prediction can be broadly categorized into feature engineering-based models, generative diffusion models, and deep learning-based models. Feature engineering-based models rely on manually designed features extracted from content, user profiles, and network structures, and establish a mapping between these features and popularity~\cite{ma2013predicting,martin2016exploring}. Generative diffusion models, inspired by epidemiological~\cite{matsubara2012rise} or stochastic processes~\cite{shen2014modeling}, have been further refined to account for specific characteristics observed in social networks, such as periodicity and user heterogeneity. Although these methods offer a certain degree of interpretability, they often struggle to capture the complex spatiotemporal patterns present in real-world cascades, thereby limiting their predictive performance.

Recently, deep learning methods, particularly those leveraging sequence models such as recurrent neural networks (RNNs)~\cite{cao2017deephawkes,gou2018learning}, Transformers~\cite{sun2023castformer}, and neural ordinary differential equations (ODEs)~\cite{cheng2024information}, have demonstrated superior performance in modeling the continuous-time dynamics and irregular event intervals observed in cascades. However, these methods typically rely solely on cascade features for popularity prediction. Notably, CasFT~\cite{jing2025casft} is the only work that attempts to use a diffusion denoising model to generate future trends. Nevertheless, these approaches still lack the ability to explicitly capture the overall trend of popularity up to the prediction time, which results in limited predictive capability. In addition, most of these methods consider only unidirectional dependencies within the cascade, further restricting their capacity to capture long-range global contextual information.

To address these challenges, we propose VNOIP, a novel method based on variational neural ODEs for information popularity prediction. First, VNOIP employs lightweight graph embeddings for both global and cascade graphs. Then, bidirectional jump ODEs with attention mechanisms are designed for enhanced sequence modeling, enabling the model to simultaneously consider information diffusion from both the past and the future. Furthermore, a variational inference module is introduced to model the uncertainty and continuous evolution of popularity trends. By jointly learning from both macroscopic popularity trends and microscopic cascade sequences, VNOIP samples the latent states of popularity dynamics from the prior and posterior distributions. In addition, a knowledge distillation loss on future latent states is employed to better align the evolution of the prior and posterior latent states. By comprehensively modeling the spatiotemporal dynamics from both cascade sequences and popularity trends, VNOIP significantly enhances the predictive capability for information popularity. 

Our contributions can be summarized as follows.
\begin{itemize}
\item A novel bidirectional jump ODEs-based model is introduced for cascade sequence modeling, enabling the capture of long-range dependencies, resulting in more expressive and informative cascade representations.

\item A variational neural ODE approach is introduced for modeling global popularity trends. By extracting features from both cascade sequences and popularity trajectories, the prior and posterior distributions of the initial state are inferred to capture trend dynamics and future uncertainty. Additionally, a knowledge distillation loss is incorporated to better align the evolution of the prior and posterior latent states.

\item Extensive experiments on benchmark datasets demonstrate that VNOIP achieves competitive performance and improved efficiency compared to SOTA baselines.
\end{itemize}

\section{Related Work}

\noindent\textbf{Feature Engineering-based Models:} Prior feature engineering based studies predict information popularity by manually extracting features from social media data. These approaches consider various factors such as content characteristics~\cite{ma2013predicting}, user attributes~\cite{martin2016exploring}, network structure~\cite{wu2016unfolding}, and temporal dynamics~\cite{yang2011patterns} to improve prediction accuracy. However, their effectiveness largely depends on the quality of manually designed features and may face limitations in generalization.

\noindent\textbf{Generative Diffusion Models:} Generative diffusion models for popularity prediction are mainly categorized into differential equation-based and stochastic process-based approaches. Differential equation-based models, often inspired by epidemiology, are adapted to account for the unique mechanisms of information diffusion, such as dynamic infection rates~\cite{matsubara2012rise} and periodic external influences~\cite{stai2018temporal}. Stochastic process-based models estimate the contribution of individual events or users, modeling mechanisms like decay, excitation, and periodicity using distributions and processes such as Poisson~\cite{shen2014modeling} and Hawkes~\cite{bao2016modeling} processes. Both approaches provide detailed representations of dynamic diffusion behaviors and improve interpretability through parameterization. However, they face challenges in capturing complex diffusion patterns and generalizing to diverse scenarios compared to deep learning methods.

\noindent\textbf{Deep Learning-based Models:} Recently, increasing attention focuses on leveraging deep learning methods for modeling the spatiotemporal characteristics of information diffusion. In early work, RNNs often serve as the backbone to model information cascades~\cite{gou2018learning,cao2017deephawkes,chen2019information}. To better capture arbitrary dependencies in long sequences, later studies incorporate attention mechanisms~\cite{sun2023castformer}. Recently, some approaches explore neural differential equations to enhance continuous-time sequence modeling~\cite{chen2018neural}. For example, Cheng et al. propose the CasDo model, which extends the discrete state transitions of RNNs to continuous-time dynamic systems~\cite{cheng2024information}. Jing et al. propose CasFT, which uses observed information cascades and dynamic cues extracted by neural ordinary differential equations as conditions, with micro-level features serving as the initial conditions for subsequent macro-level spatiotemporal pattern prediction~\cite{jing2025casft}.
Moreover, modeling the uncertainty of macro-level diffusion significantly affects predictive performance. Common approaches introduce deep generative models such as normalizing flows~\cite{xu2021casflow} and denoising diffusion models~\cite{cheng2024information,jing2025casft} to capture uncertainty in prediction results and improve accuracy.

\begin{figure*}[ht]
    \centering
    \includegraphics[width=1.\linewidth]{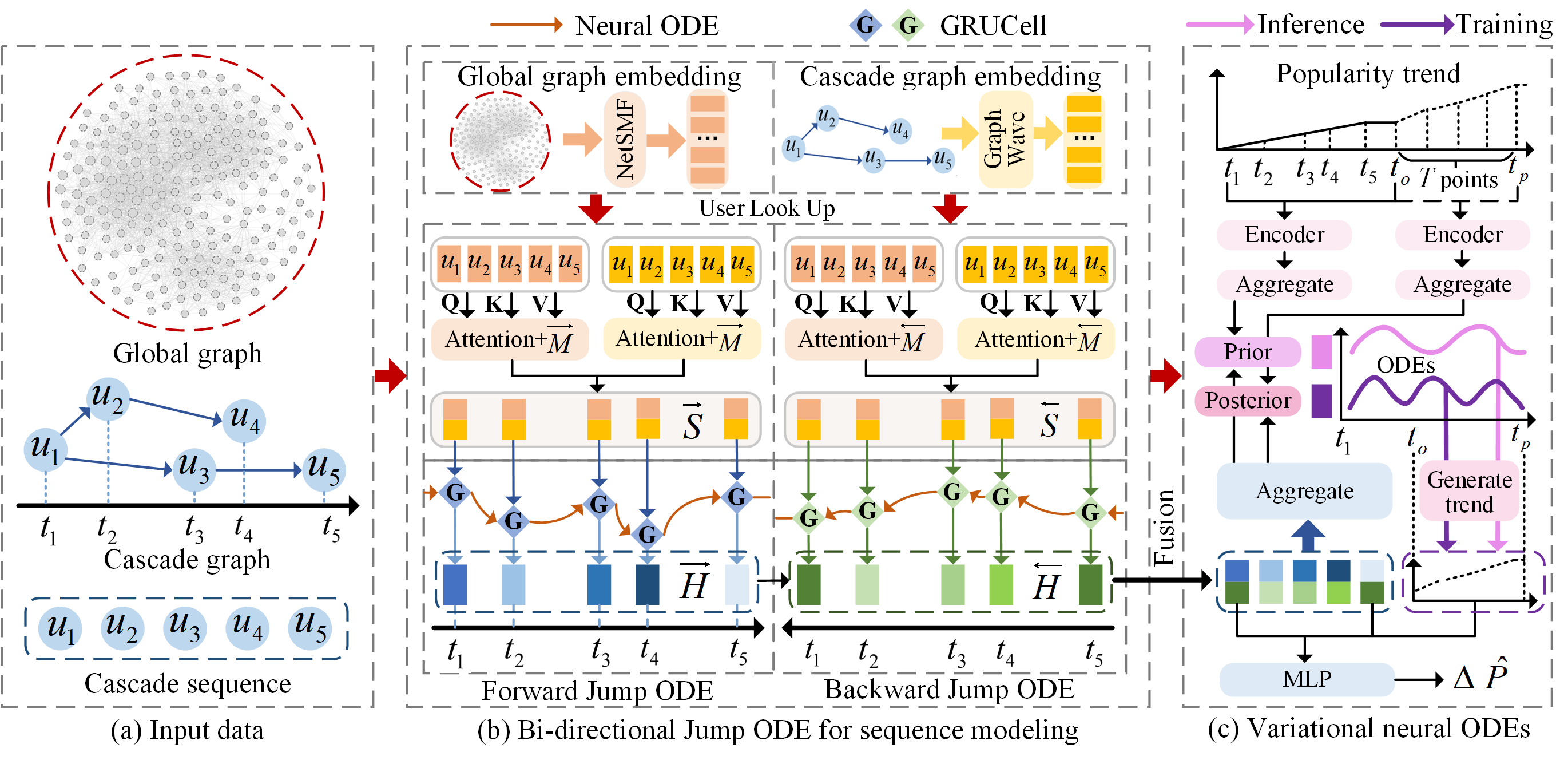}
    \caption{The overview of VNOIP. (a) The global graph, the observed cascade graph, and the cascade sequence. (b) Bi-directional Jump ODEs with attention to enhance the representation capability of sequence context. (c) Variational neural ODEs model the overall trend by jointly aggregating cascade patterns and popularity trend features.
} 
    \label{overview}
\end{figure*}

\section{Problem Statement}

\noindent\textbf{Global Graph:} The global graph $\mathcal{G}^g =
\{\mathcal{V}^g, \mathcal{E}^g\}$ represents a social/citation network. The nodes are users and the edges are relationships between them.

\noindent\textbf{Cascade Graph:} Given an information item and the observed record of its diffusion, the cascade graph is defined as $\mathcal{G}^c = \{\mathcal{V}^c, \mathcal{E}^c\}$, $\mathcal{V}^c = \{u_0, u_1, u_2, \ldots, u_n\}$ denotes the set of users involved in spreading the information, while $\mathcal{E}^c = \{e_1, e_2, \ldots, e_m\}$ specifies the connections among these users. Each edge $e_i = (u_j, u_i, t_i)$ indicates that user $u_i$ has reposted or forwarded the information from user $u_j$ at timestamp $t_i$.

\noindent\textbf{Cascade Sequence:} Let $\mathcal{C}(t_o)$ be a collection of users arranged based on their timestamps until the observation time $t_o$, i.e., $\mathcal{C}(t_o) = \{(u_0, t_0), (u_1, t_1), \ldots, (u_n, t_n)\}$, where $t_0$$<$$t_1$$<$$\cdots$$<$$t_n$$\leq t_o$.

\noindent\textbf{Information Cascade Popularity Prediction:} 
Given a cascade $c$ observed up to time $t_o$, together with the global graph $\mathcal{G}^g$, the observed cascade graph $\mathcal{G}^c(t_o)$, and the cascade sequence $\mathcal{C}(t_o)$, the objective of popularity prediction is to estimate the incremental popularity $\Delta P = P(t_p) - P(t_o)$, where $P(t_i)$ denotes the popularity at time $t_i$, $t_p \gg t_o$, and $t_p$ is the prediction horizon.

\section{Methodology}
\subsection{Overview}
Fig.~\ref{overview}(a) shows the input data, including the global graph, the cascade graph, and the cascade sequence.
In Fig.~\ref{overview}(b), VNOIP models the cascade sequence using bidirectional jump ODEs based on both global and cascade graph embeddings, which capture both past and future influence dynamics in the cascade process.
Fig.~\ref{overview}(c) presents the prediction module based on variational neural ODEs. The aggregated spatiotemporal patterns from the bidirectional jump ODEs and the popularity trend are encoded into latent variables through variational inference. The model then generates the popularity trend from the learned latent dynamics using neural ODEs. Finally, the generated trend and cascade hidden states are fed into MLPs to predict the incremental trend.

\subsection{Global and Cascade Graph Embeddings}
Since information popularity prediction targets macro-level patterns, the underlying networks are often large in scale. Consequently, representing network structures directly with advanced graph neural networks becomes challenging. To address this, lightweight graph embedding techniques are adopted to extract distinctive structural patterns from both the global graph and the information cascade graph, as suggested in previous studies~\cite{xu2021casflow,jing2025casft}. For the large-scale global graph, NetSMF~\cite{qiu2019netsmf} is employed due to its efficiency in handling large networks and its ability to quickly learn user interactions, embedding global structural features into node representations $\mathbf{X}^g$. For the cascade graph, GraphWave~\cite{donnat2018learning} is utilized to effectively capture local structural patterns and obtain the node embeddings $\mathbf{X}^c$.

\subsection{Bi-directional Jump ODEs for Sequence Modeling}
Most previous studies model information cascades by considering only the unidirectional process from the initial time to the final time. However, such models can only capture partial dependencies within the sequence, thereby limiting their capacity for comprehensive context representation. Therefore, this paper proposes bidirectional jump ODEs with attention mechanisms, which aggregate both past and future information at each position in the information cascade. This approach alleviates the bias caused by strict sequential order and enhances the expressive power of the representation.

\subsubsection{Bi-directional Context Representation}
To enhance the model’s ability to capture long-range dependencies, the attention mechanism is utilized to obtain the context representation. Specifically, following previous work~\cite{jing2025casft}, two self-attention layers are independently applied to both the cascade graph and the global graph views.

For the global graph, the user embeddings $S^g = [X^g_{u_0}, \ldots, X^g_{u_n}]$ are retrieved according to the sequence of users involved in the information cascade. The scaled dot-product attention~\cite{attention} is then applied to obtain bidirectional contextual representations for the sequence as follows.
\begin{equation}
\begin{split}
&\mathop{S^g}\limits ^{\rightarrow}=\text{Attention}(S^g, S^g, S^g, \mathop{M}\limits ^{\rightarrow}),\\
&\mathop{S^g}\limits ^{\leftarrow}=\text{Attention}(S^g, S^g, S^g, \mathop{M}\limits ^{\leftarrow}),\\
&\text{Attention}(Q, K, V, M)=\text{Softmax}\left(\frac{QK^\top}{\sqrt{d}} + M\right)V,
\end{split}
\label{attention}
\end{equation}
where $\mathop{S^g}\limits ^{\rightarrow}$ and $\mathop{S^g}\limits ^{\leftarrow}$ denote the bidirectional contextual representations, $\mathop{M}\limits ^{\rightarrow}$ and $\mathop{M}\limits ^{\leftarrow}$ are the forward and backward mask matrices (i.e., past-to-future and future-to-past masks).

For the cascade graph, the sequence embeddings are defined as $S^c = [X^c_{u_0} + E(t_0), \ldots, X^c_{u_n} + E(t_n)]$, where $X^c_{u_i}$ is user $u_i$' cascade graph embedding and $E(t_i)$ is the temporal embedding that encodes the specific position of user $u_i$ in the cascade sequence. The temporal encoding $E(t_i)$ is defined as follows.
\begin{equation}
E(t_i)_j =
\begin{cases}
\cos\left( \dfrac{t_i}{10000^{\frac{j-1}{d}}} \right), & \text{if } j \text{ is odd}, \\
\sin\left( \dfrac{t_i}{10000^{\frac{j}{d}}} \right), & \text{if } j \text{ is even},
\end{cases}
\end{equation}
where $t_i$ denotes the timestamp of the $i$-th user in the sequence, $j$ is the $j$-th dimension, and $d$ is the embedding dimension. Then, the bidirectional contextual representations for the cascade graph, $\mathop{S^c}\limits ^{\rightarrow}$ and $\mathop{S^c}\limits ^{\leftarrow}$, can then be obtained in the same manner as in Eq.~(\ref{attention}).

Finally, the forward and backward representations from both views are concatenated to form the final contextual representations, i.e., $\mathop{S}\limits ^{\rightarrow} = [\mathop{S^c}\limits ^{\rightarrow}; \mathop{S^g}\limits ^{\rightarrow}]$ and $\mathop{S}\limits ^{\leftarrow} = [\mathop{S^c}\limits ^{\leftarrow}; \mathop{S^g}\limits ^{\leftarrow}]$, where $[\cdot;\cdot]$ denotes vector concatenation.

\subsubsection{Jump ODEs for Sequence Modeling}
First, the initial state of a cascade sequence is initialized as a trainable vector $H$. Then, self-gating is applied to generate separate forward and backward initial states $ \mathop{H}\limits ^{\rightarrow}(t_0)$ and $\mathop{H}\limits ^{\leftarrow}(t_n)$ for the jump ODEs, respectively.
\begin{equation}
\begin{split}
    \mathop{H}\limits ^{\rightarrow}(t_0) = H \odot \text{sigmoid}(W^fH+b^f),\\
    \mathop{H}\limits ^{\leftarrow}(t_n) = H \odot \text{sigmoid}(W^bH+b^b),
\end{split} 
\end{equation}
where $\odot$ denotes element-wise multiplication, and $W$ and $b$ are learnable parameters.

Then, the continuous-time evolution of the hidden states $\mathop{H}\limits ^{\rightarrow}(t)$ and $\mathop{H}\limits ^{\leftarrow}(t)$ over the intervals $[t_i^+, t_{i+1}]$ and $[t_{i-1}, t_i^-]$ is modeled by neural ODEs, as shown in Eq.~(\ref{ode1}).
\begin{equation}
\begin{split}
\frac{\mathrm{d}\mathop{H}\limits ^{\rightarrow}(t)}{\mathrm{d}t} = f_{o_1}(\mathop{H}\limits ^{\rightarrow}(t)),\quad
\frac{\mathrm{d}\mathop{H}\limits ^{\leftarrow}(t)}{\mathrm{d}t} = f_{o_2}(\mathop{H}\limits ^{\leftarrow}(t)),
\end{split}
\label{ode1}
\end{equation}
where $f_{o_1}(\cdot)$ and $f_{o_2}(\cdot)$ are multi-layer perceptrons (MLPs).

When an action occurs at time $t_i$, the states of the cascade sequence, $\mathop{H}\limits ^{\rightarrow}(t_i)$ and $\mathop{H}\limits ^{\leftarrow}(t_i)$, are updated via a jump process using GRUs, as shown in Eq.~(\ref{gru}).
\begin{equation}
\begin{split}
    \mathop{H}\limits ^{\rightarrow}(t_i^+) = \text{GRUCell}(\mathop{H}\limits ^{\rightarrow}(t_i), \mathop{S_i}\limits ^{\rightarrow}),\\
     \mathop{H}\limits ^{\leftarrow}(t_i^-) = \text{GRUCell}(\mathop{H}\limits ^{\leftarrow}(t_i), \mathop{S_i}\limits ^{\leftarrow}),
\end{split}
\label{gru}
\end{equation}
where $\mathop{S_i}\limits ^{\rightarrow}$ and $\mathop{S_i}\limits ^{\leftarrow}$ denote the $i$-th positional contextual representations obtained from the attention layers, which are used to enhance the receptive field of sequence modeling.

\subsubsection{Feature Fusion}
By employing both forward and backward jump ODEs, we obtain the cascade sequence states $\mathop{H}\limits ^{\rightarrow} = [\mathop{H}\limits ^{\rightarrow}(t_0^+), \ldots, \mathop{H}\limits ^{\rightarrow}(t_n^+)]$ and $\mathop{H}\limits ^{\leftarrow} = [\mathop{H}\limits ^{\leftarrow}(t_0^-), \ldots, \mathop{H}\limits ^{\leftarrow}(t_n^-)]$, respectively. Finally, a channel attention fusion mechanism is applied to produce the final sequence state $\bar{H}$ as follows.
\begin{equation}
\begin{split}
&\gamma(t_i) = \frac{\exp(a \cdot W^a H(t_i))}{\sum_{H^{'}(t_i) \in \{ \mathop{H}\limits ^{\rightarrow}(t_i^+), \mathop{H}\limits ^{\leftarrow}(t_i^-)\} } \exp(a \cdot W^a H^{'}(t_i))},
\\
&\bar{H}(t_i) = \sum_{H^{'}(t_i) \in \{ \mathop{H}\limits ^{\rightarrow}(t_i^+), \mathop{H}\limits ^{\leftarrow}(t_i^-)\} } \gamma^{'}(t_i) H^{'}(t_i),
\end{split}
\label{fusion}
\end{equation}
where $a$ and $W^a$ are learnable parameters. The final sequence representation $\bar{H}(t_i)$ is computed as the weighted sum of the forward and backward hidden states at each time step. To further improve training stability, Layer Normalization is applied to the output, i.e., $\bar{H} = \text{LayerNorm}(\bar{H})$.

\subsection{Variational Encoder for Popularity Trend}
To generate the complete popularity trend and facilitate the final prediction, the interval $[t_o, t_p]$ is evenly divided into $T$ points, and the popularity at these points is used during training. Furthermore, a variational neural ODEs module is employed to model the dynamics of popularity evolution.
During the training phase, the popularity trajectories $P^{\text{ctx}}$ and $P^{\text{tgt}}$ are constructed. Specifically, $P^{\text{ctx}} = [P(t_0),\ldots,P(t_o)]$ represents the popularity in the cascade observed up to time $t_o$. Building upon $P^{\text{ctx}}$, $P^{\text{tgt}} = [P^{\text{ctx}}, P(t_o + \frac{(t_p - t_o)}{T}), P(t_o + 2\frac{(t_p - t_o)}{T}), \ldots, P(t_p)]$ incorporates the cumulative popularity at $L$ additional future time points, denoted as $P^{\text{tgt}}(t)$ for $t_o < t \leq t_p$, where $T$ is a hyperparameter.

Subsequently, following the framework of variational inference, the Evidence Lower BOund (ELBO) can be formulated as Eq.~(\ref{elbo}).
\begin{equation}
\begin{split}
    \log p(P^{\text{tgt}} | P^{\text{ctx}}, \bar{H}) &\geq 
    \mathbb{E}_{q(z_0 | P^{\text{tgt}}, \bar{H})} \big[ \log p(P^{\text{tgt}} | z_0, \bar{H}) \big] 
    \\
    &- \mathrm{KL}\big(q(z_0 | P^{\text{tgt}}, \bar{H}) \,\|\, p(z_0 | P^{\text{ctx}}, \bar{H})\big),
\end{split}
\label{elbo}
\end{equation}
\subsubsection{Latent Variables of Initial State}
To account for the influence of user structure and event popularity on spatiotemporal features, both the popularity trajectory and the cascade sequence are utilized as two views to encode the latent variable. Specifically, for the popularity sequences $P^{\text{ctx}}$ and $P^{\text{tgt}}$, MLPs are employed to obtain $Z^{\text{ctx}}$ and $Z^{\text{tgt}}$, $Z_i = f([P(t_i); t_i])$, where $Z_i$ denotes the vector representation at the $i$-th position, $[P(t_i); t_i]$ denotes the concatenation the popularity at time $t_i$ with the timestamp $t_i$. 
Then, a trainable vector $Z^{p}$ is utilized as the query to aggregate the features $Z^{\text{ctx}}$ and $Z^{\text{tgt}}$ through attention layers, as illustrated below.
\begin{equation}
\begin{split}
    \bar{Z}^{\text{ctx}} &= \text{Attention}(Z^{p}, Z^{\text{ctx}}, Z^{\text{ctx}}),\\
    \bar{Z}^{\text{tgt}} &= \text{Attention}(Z^{p}, Z^{\text{tgt}}, Z^{\text{tgt}}),
\end{split}
\end{equation}
In the similar way, a trainable vector $Z^{h}$ is utilized as the query to aggregate the cascade hidden states $\bar{H}$ through an attention layer.
\begin{equation}
    \bar{Z}^{h} = \text{Attention}(Z^{h}, \bar{H}, \bar{H}).
\end{equation}
By concatenating the aggregated feature ($\bar{Z}^{\text{ctx}}$ or $\bar{Z}^{\text{tgt}}$) from the popularity trend and the aggregated feature $\bar{Z}^{h}$ from the cascade sequence, the initial representation incorporates both the macroscopic and microscopic spatiotemporal characteristics. Subsequently, MLPs are employed to map these features to Gaussian distributions, and the latent variable $z$ for modeling the popularity trend is obtained via the reparameterization trick. 
\begin{equation}
\begin{split}
    z^{\text{prior}}(t_0) &\sim N\left(\mu_p([\bar{Z}^{\text{ctx}};\bar{Z}^{h}]),\, \sigma_p([\bar{Z}^{\text{ctx}};\bar{Z}^{h}])\right),\\
    z^{\text{post}}(t_0) &\sim N\left(\mu_q([\bar{Z}^{\text{tgt}};\bar{Z}^{h}]),\, \sigma_q([\bar{Z}^{\text{tgt}};\bar{Z}^{h}])\right),
\end{split}
\end{equation}
where $\mu_p(\cdot)$, $\mu_q(\cdot)$, $\sigma_p(\cdot)$, and $\sigma_q(\cdot)$ can be learned from MLPs. To ensure that $\sigma(Z) > 0$, we parameterize the standard deviation as $\sigma(Z) = 0.1 + 0.9 \cdot \mathrm{sigmoid}(\mathrm{MLP}(Z))$.

\subsubsection{Neural ODEs for Popularity Trend Generation}
After obtaining the initial value $z(t_0)$, neural ODEs are employed to generate the complete popularity trajectory. Meanwhile, to model the uncertainty in the generation process and to ensure that the instantaneous increments are non-negative, the expected value of a truncated normal distribution is adopted for the popularity increment $\mathrm{d}P(t)/\mathrm{d}t$. The specific generation process is as follows.
\begin{equation}
\begin{split}
    \frac{\mathrm{d}z(t)}{\mathrm{d}t} &= f_{z}\left(z(t)\right), \\
    \frac{\mathrm{d}P(t)}{\mathrm{d}t} &= \mu_f\left(z(t)\right) + \sigma_f\left(z(t)\right) \cdot \frac{\phi(t)}{1 - \Phi(t)}, \\
    \phi(t) &= \frac{1}{\sqrt{2\pi}} \exp\left( -\frac{\alpha(t)^2}{2} \right), \\
    \Phi(t) &= \frac{1}{2}\left[1 + \operatorname{erf}\left(\frac{\alpha(t)}{\sqrt{2}}\right)\right], 
    \alpha(t) = \frac{ -\mu_f\left(z(t)\right) }{ \sigma_f\left(z(t)\right) }
\end{split}
\end{equation}
where $f_{z}(\cdot)$, $\mu_f(\cdot)$, and $\sigma_f(\cdot)$ are learned by MLPs, $\operatorname{erf(\cdot)}$ is the error function.

Given the initial prior $z^{\text{prior}}(t_0)$ and posterior $z^{\text{post}}(t_0)$, and employing a shared neural ODE, an ODE solver computes the future trends $P^{\text{prior}}$ and $P^{\text{post}}$ for both the prior and posterior. In addition, the latent trajectories $z^{\text{prior}} = [z^{\text{prior}}(t_0), \ldots, z^{\text{prior}}(t_p)]$ and $z^{\text{post}} = [z^{\text{post}}(t_0), \ldots, z^{\text{post}}(t_p)]$ are obtained.

\subsection{Knowledge Distillation of Future Latent States}
Although the prior and posterior latent variables are evolved using a shared neural ODE, slight differences in their initial states can result in substantial divergence after long-term dynamics. To mitigate this effect and promote similarity between the prior and posterior latent variables over time, a KL divergence-based knowledge distillation loss is applied at the final time step to regularize the evolved latent variables, which is shown as follows.
\begin{equation}
\begin{split}
    \mathcal{L}_{\mathrm{kd}} = \frac{1}{2} \Bigg(
        &\,\mathrm{KL}\left(z^{\mathrm{prior}}(t_p) \;\middle\|\; z^{\mathrm{post}}(t_p)\right) +\\
        &\,\mathrm{KL}\left(z^{\mathrm{post}}(t_p) \;\middle\|\; z^{\mathrm{prior}}(t_p)\right)
    \Bigg)
\end{split}
\end{equation}
This symmetric formulation penalizes discrepancies in both directions and thus more effectively aligns the prior and posterior latent representations at time $t_p$.

\subsection{Model Training and Popularity Prediction}
During training, the final prediction is obtained via a decoder as Eq.~(\ref{decoder1}).
\begin{equation}
    \Delta\hat{P} = \operatorname{Softplus}\big(f_d([\bar{H}(t_0); \bar{H}(t_n); P^{\text{post}}])\big),
\label{decoder1}
\end{equation} 
where $f_d$ denotes MLPs and $P^{\text{post}}$ indicates the generated future trend from posterior latent variable. The main loss is computed by the mean squared logarithmic error (MSLE).
\begin{equation}
    \mathcal{L}_{\mathrm{main}} = \left( 
        \log_2 \Big( \Delta P + 1 \Big) 
        - 
        \log_2 \left( \Delta \hat{P} + 1 \right)
    \right)^2.
\end{equation}
To encourage the generated trends to better match the ground truth, we introduce a regression loss $\mathcal{L}_{\mathrm{rg}}$ to approximate the reconstruction loss, defined as the MSLE between the generated trends (i.e., $P^{\mathrm{prior}}$ and $P^{\mathrm{post}}$) and the ground-truth trend $P$ over the interval $[t_o, t_p]$ at $T$ time points.

The overall loss function for optimization is given by
\begin{equation}
\begin{split}
\mathcal{L}_{\mathrm{total}}&=\mathcal{L}_{\mathrm{main}}+\lambda_1\mathcal{L}_{\mathrm{rg}}+\lambda_2(\mathcal{L}_{\mathrm{kl}} + \mathcal{L}_{\mathrm{kd}}),\\
\mathcal{L}_{\mathrm{kl}}&=\mathrm{KL}\big(q(z^{\mathrm{post}}(t_0) ) \,\|\, p(z^{\mathrm{prior}}(t_0)\big),
\end{split}
\end{equation}
where $\lambda_1$ and $\lambda_2$ are hyperparameters that balance the contributions of each loss term, and the prior and posterior distributions are defined in Eq.~(10).

After training is completed via backpropagation and optimization, during the testing phase, the trend generated from the prior latent variable $z^{\mathrm{prior}}(t_0)$ is utilized to assist in predicting the incremental popularity, as shown below.
\begin{equation}
    \Delta\hat{P} = \operatorname{Softplus}\big(f_d([\bar{H}(t_0); \bar{H}(t_n); P^{\text{prior}}])\big).
\label{decoder}
\end{equation}

\section{Experiments}

\subsection{Datasets}
We evaluate VNOIP and other baselines on three real-world datasets: Twitter, APS~\cite{shen2014modeling}, and Weibo~\cite{cao2017deephawkes}.
\textit{Twitter} consists of tweet cascades collected between March 1 and April 15, 2022, based on hashtags.
\textit{APS} contains citation cascades from papers published in American Physical Society journals between 1893 and 1997.
\textit{Weibo} includes repost cascades from China’s largest microblogging platform.
For each dataset, we follow previous works~\cite{jing2025casft} to set the observation and prediction periods. Specifically, observation windows are 1 and 2 days for Twitter, 3 and 5 years for APS, and 0.5 and 1 hour for Weibo. Prediction periods are set to 15 days, 20 years, and 24 hours, respectively. Cascades with fewer than 10 participants during the observation window are excluded. We split each dataset into 70\% for training, 15\% for validation, and 15\% for testing. More details are provided in Table 1.

\begin{table}[htbp]
\centering
\resizebox{\columnwidth}{!}{%
    \begin{tabular}{lccc}
    \hline
    \textbf{Dataset} & \textbf{Twitter} & \textbf{APS} & \textbf{Weibo} \\
    \hline
    Cascades & 86,764 & 207,685 & 119,313 \\
    Nodes in $\mathcal{G}^g$ & 490,474 & 616,316 & 6,738,040\\
    Edges in $\mathcal{G}^g$ & 1,903,230 & 3,304,400 & 15,249,636 \\
    Avg. popularity & 94 & 51 & 240 \\
    \hline
    \multicolumn{4}{l}{\textit{Number of cascades in two observation settings}} \\
    Train (1d/3y/0.5h) & 7,308 & 18,511 & 21,463 \\
    Val (1d/3y/0.5h)   & 1,566 & 3,967 & 4,599 \\
    Test (1d/3y/0.5h)  & 1,566 & 3,966 & 4,599 \\
    Train (2d/5y/1h)   & 10,983 & 32,102 & 29,908 \\
    Val (2d/5y/1h)     & 2,353 & 6,879 & 6,409 \\
    Test (2d/5y/1h)    & 2,353 & 6,879 & 6,408 \\
    \hline
    \end{tabular}%
}
\caption{Statistics of the three datasets.}
\end{table}

\subsection{Baselines}
The baselines used for comparison include the following representative methods:
\textbf{DeepCas}~\cite{li2017deepcas} utilizes a bi-directional GRU with attention for effective sequence modeling. \textbf{DeepHawkes}~\cite{cao2017deephawkes} combines the Hawkes process with deep learning to capture user influence, self-excitation, and temporal decay effects in cascade dynamics. \textbf{VaCas}~\cite{zhou2020variational} introduces a hierarchical graph learning framework and leverages a variational autoencoder to model uncertainty. \textbf{CasFlow}~\cite{xu2021casflow} considers both local and global graph structures to characterize user behavior for popularity prediction. \textbf{CTCP}~\cite{lu2023continuous} proposes an evolution learning module that updates user and cascade states in real time as diffusion unfolds. \textbf{CasDO}~\cite{cheng2024information} proposes a novel framework that integrates probabilistic diffusion models with neural ODEs for information cascade popularity prediction. \textbf{CasFT}~\cite{jing2025casft} employs probabilistic diffusion models to dynamically generate cues for modeling popularity trends.

\subsection{Experimental Settings}
In our experiments, the batch size is set to 100. The hyperparameters ($\lambda_1$, $\lambda_2$, $T$) are tuned to achieve the best performance on the validation set (details can be found in the section on Sensitivity Analysis of Hyperparameters), and early stopping with a patience of 15 epochs is applied. The dimensions of both the cascade graph and global graph representations are set to 40. The functions $f_{o_1}$ and $f_{o_2}$ are implemented as three-layer MLPs with \texttt{Softplus} activation functions. The functions $f_{z}(\cdot)$, $\mu_{f}(\cdot)$, and $\sigma_{f}(\cdot)$ are also implemented as three-layer MLPs with \texttt{Tanh} activation functions. All other MLP modules are two-layer networks with \texttt{ReLU} activations. The dimensions of hidden states and latent variables are set to 64, except for the Twitter dataset with a 2-day observation window and the Weibo dataset, where the dimension is set to 128. The model parameters are optimized using the Adam optimizer with a learning rate of $2 \times 10^{-3}$. For sequence modeling, the ODE solver is set to \texttt{euler}, while for popularity generation, it is set to \texttt{dopri5}.

\subsection{Evaluation Metrics}
Two commonly used metrics, mean squared logarithmic error (MSLE) and mean absolute percentage error (MAPE), are employed to evaluate the performance of models:

\begin{equation}
\begin{split}
    &\text{MSLE} = \frac{1}{B} \sum_{b=1}^{B} \left( \log_2(\Delta P_b + 1) - \log_2(\Delta\hat{P}_b + 1) \right)^2,
    \\
&\text{MAPE} = \frac{1}{B} \sum_{b=1}^{B} \left| \frac{\log_2(\Delta P_b + 2) - \log_2(\Delta \hat{P}_b + 2)}{\log_2(\Delta P_b + 2)} \right|,
\end{split}
\end{equation}
where $\Delta P_b$ denotes the true incremental popularity of $b$-th cascade, $\Delta \hat{P}_b$ represents the predicted incremental popularity, and $B$ is the total number of test cascades.

\subsection{Experimental Results}

\begin{table*}[htbp]
\centering


\setlength{\tabcolsep}{1mm}

\renewcommand{\arraystretch}{1.1}
\begin{tabular}{c|cc|cc|cc|cc|cc|cc}
\hline
\multirow{3}{*}{\textbf{Method}} 
    & \multicolumn{4}{c|}{\textbf{Twitter}} 
    & \multicolumn{4}{c|}{\textbf{APS}} 
    & \multicolumn{4}{c}{\textbf{Weibo}} \\
\cline{2-13}
    & \multicolumn{2}{c|}{1-day} & \multicolumn{2}{c|}{2-days} 
    & \multicolumn{2}{c|}{3-years} & \multicolumn{2}{c|}{5-years} 
    & \multicolumn{2}{c|}{0.5-hour} & \multicolumn{2}{c}{1-hour} \\
    & MSLE & MAPE & MSLE & MAPE 
    & MSLE & MAPE & MSLE & MAPE 
    & MSLE & MAPE & MSLE & MAPE \\
\hline                    
DeepCas       & 6.330 & 0.657 & 5.715 & 0.667 & 2.105 & 0.287 & 1.926 & 0.346  & 4.646 & 0.326 & 3.553 & 0.353  \\
DeepHawkes    & 5.934 & 0.502 & 4.849 & 0.519 & 1.914 & 0.282 & 1.815 & 0.337  & 2.874 & 0.304 & 2.743 & 0.335 \\
VaCas         & 5.512 & 0.479 & 4.214 & 0.487 & 1.776 & 0.269 & 1.695 & 0.301  & 2.525 & 0.285 & 2.345 & 0.299 \\
CasFlow       & 4.779 & 0.415 & 3.688 & 0.422 & 1.437 & 0.240 & 1.335 & 0.262  & 2.337 & 0.266 & 2.223 & 0.294 \\
CTCP          & 5.399 & 0.376 & 3.601 & 0.377 & 1.767 & 0.305 & 1.375 & 0.291  & 2.557 & 0.306 & 2.297 & 0.301 \\
CasDo         & 4.636 & 0.425 & 3.714 & 0.397 & \textbf{1.192} & \textbf{0.217} & \underline{1.165} & \textbf{0.222}  & \underline{2.175} & \underline{0.254} & \underline{2.141} & \textbf{0.256} \\
CasFT         & \underline{4.146} & \underline{0.391} & \underline{3.298} & \underline{0.367} & 1.301 & 0.234 & 1.188 & 0.251 & 2.261 & 0.268 & 2.175 & 0.289 \\
\hline
VNOIP-B       & 4.144 & 0.381 & 3.152 & 0.359 & 1.251 & 0.229 & 1.183 & 0.253 & 2.209 & 0.252 & 2.085 &  0.273 \\
VNOIP-F       & 4.232 & 0.408 & 3.382 & 0.443 & 1.282 & 0.225 & 1.267 & 0.266  &  2.235 & 0.263 & 2.142 &  0.276 \\
VNOIP-V       & 4.195 & 0.389 & 3.269 & 0.369 & 1.275 & 0.228 & 1.190 & 0.256 & 2.216 & 0.259 & 2.069 & 0.270 \\
VNOIP-K      & 4.180 & 0.382 & 3.259 & 0.414 & 1.237 & 0.224 & 1.185 & 0.253 & 2.174 & 0.255 & 2.111 & 0.282  \\
VNOIP         & \textbf{4.026} & \textbf{0.376} & \textbf{3.061} & \textbf{0.351} & \underline{1.247} & \underline{0.222} & \textbf{1.159} & \underline{0.247} & \textbf{2.139} & \textbf{0.251} & \textbf{2.037} & \underline{0.268} \\
\hline
\end{tabular}
\caption{Performance comparison between baselines and VNOIP on three datasets across different observation times measured by MSLE and MAPE (lower is better).}
\label{result}
\end{table*}

Table~\ref{result} presents the performance comparison of different methods on the Twitter, APS, and Weibo datasets. On the Twitter dataset, VNOIP demonstrates notable improvements over the baselines for both the 1-day and 2-day observation windows. For the 1-day window, VNOIP achieves a 2.9\% reduction in MSLE and a 3.8\% reduction in MAPE compared to the second-best method, CasFT. For the 2-day window, VNOIP yields a 7.2\% reduction in MSLE and a 4.4\% reduction in MAPE over CasFT. On the APS dataset, CasDo achieves better results than VNOIP for the 3-year window, which may be due to its explicit modeling of structural uncertainty. VNOIP, however, still achieves results that are close to those of CasDo, demonstrating its strong predictive capability even without explicitly modeling structural uncertainty. For the 5-year window, VNOIP slightly outperforms CasDo in MSLE, while CasDo maintains the best MAPE. On the Weibo dataset, VNOIP demonstrates strong overall performance. In particular, for the 1-hour window, VNOIP achieves a 4.9\% reduction in MSLE compared to CasDo. These results demonstrate that VNOIP is highly competitive, and consistently outperforming most existing baselines.

\subsection{Ablation Study}
We conduct ablation studies on VNOIP with the following configurations.
\textbf{VNOIP-B}: Utilizes only the forward jump ODE for sequence modeling to evaluate the impact of bidirectional modeling.
\textbf{VNOIP-F}: Removes the popularity trend modeling and directly predicts via bidirectional sequence features.
\textbf{VNOIP-V}: Disables variational inference, using only the generation results from $z^{\mathrm{ctx}}$.
\textbf{VNOIP-K}: Removes the knowledge distillation loss. The results in Table~\ref{result} show that, for most datasets, excluding the modeling of overall trends leads to a substantial decline in prediction performance. The second most notable drop occurs when variational encoding is omitted. These results suggest that VNOIP’s approach of leveraging variational neural ODEs for overall trend modeling is indeed effective.

\subsection{Efficiency of VNOIP}
\begin{figure}[ht]
    \flushleft
    \includegraphics[width=1\linewidth]{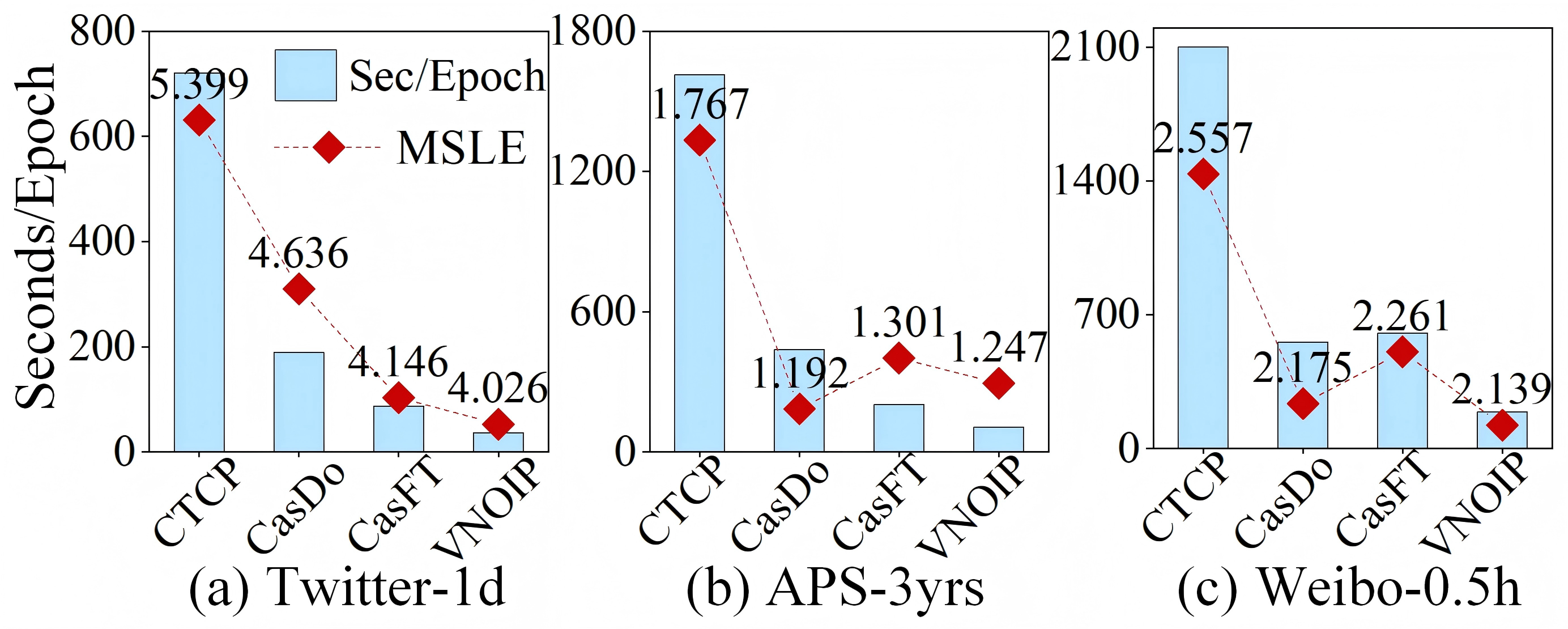}
    \caption{Comparison of the performance and training time.}
    \label{time}
\end{figure}

To demonstrate the superior efficiency of VNOIP in model training, we compare the training times of recent SOTA methods on three datasets. As shown in Fig.~(\ref{time}), VNOIP achieves a significant improvement in efficiency. Unlike CasDo and CasFT, VNOIP does not employ the more complex denoising diffusion probabilistic models to capture process uncertainty. The lower efficiency of CTCP is primarily due to its use of multiple LSTMs to model the dynamic cascade graph, which increases computational overhead.

\subsection{Sensitive Analysis of Hyperparameter}
\subsubsection{Effect of Weights $\lambda_1$ and $\lambda_2$}
\begin{figure}[ht]
    \centering
    \includegraphics[width=1.\linewidth]{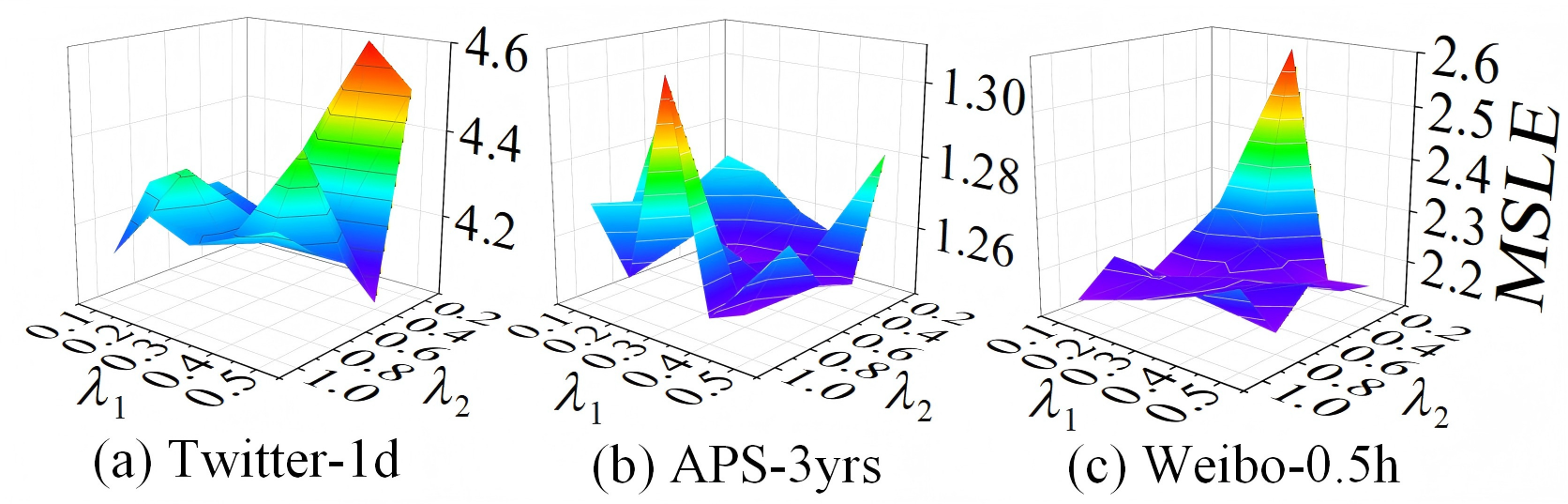}
    \caption{Effect of weights $\lambda_1$ and $\lambda_2$.}
    \label{weight}
\end{figure}

To demonstrate the effect of the weights $\lambda_1$ and $\lambda_2$ on prediction performance, we train models with different combinations of $\lambda_1=[0.1, 0.2, 0.3, 0.4, 0.5]$ and $\lambda_2=[0.2, 0.4, 0.6, 0.8, 1.0]$ on Twitter-1day, APS-3years, and Weibo-0.5hour. The results are shown in the Fig.~(\ref{weight}). The results show that there exist optimal regions in the $\lambda_1$–$\lambda_2$ space where the MSLE achieves its minimum, suggesting that a proper balance between the corresponding loss terms is crucial for achieving the best prediction accuracy. 

\subsubsection{Effect of Number of Future Data Points $T$}
\begin{figure}[ht]
    \centering
    \includegraphics[width=1.\linewidth]{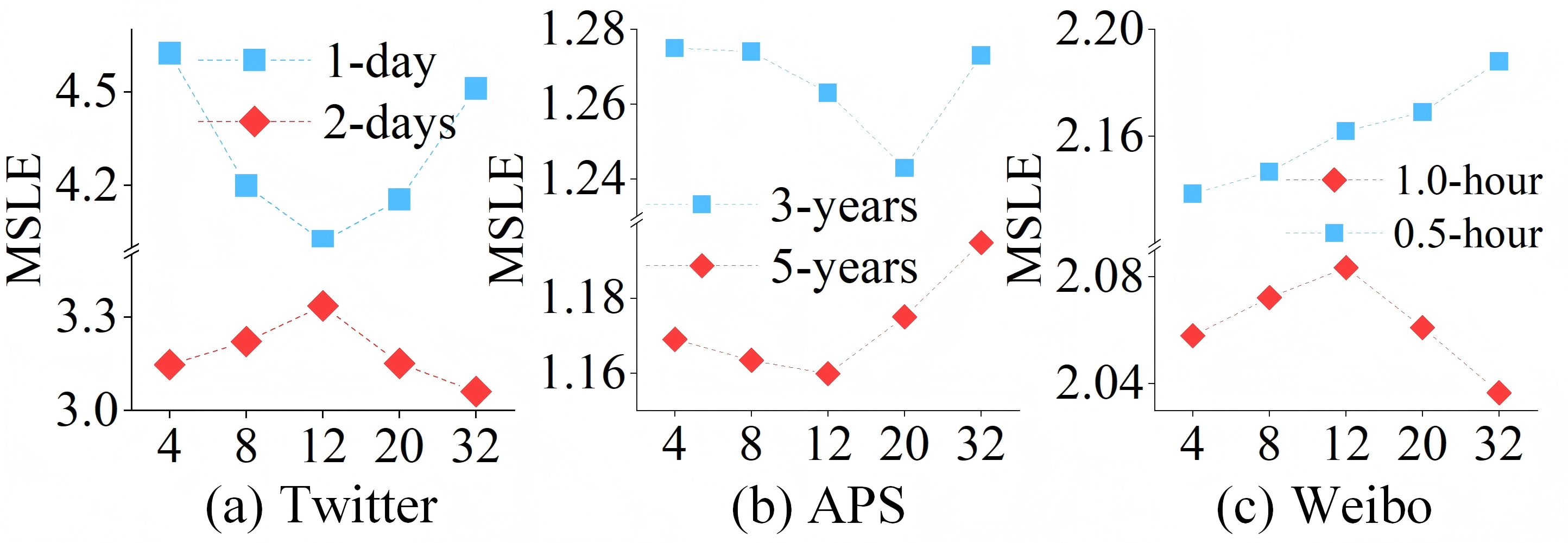}
    \caption{Effect of number of future data points $T$.}
    \label{window}
\end{figure}

To investigate the impact of the hyperparameter $T$ on prediction performance, we train models on each dataset using various values of $T= [4, 8, 12, 20, 32]$. The experimental results in Fig.~(\ref{window}) indicate that, in most cases, partitioning future trends into finer intervals does not necessarily improve prediction accuracy. This could be attributed to instances of early saturation, where the growth halts prematurely, leading to zero increments within certain intervals and thereby increasing the difficulty of model training.

\section{Conclusion}
This paper proposes VNOIP, a novel variational neural ODE-based approach for information popularity prediction. VNOIP jointly models cascade sequences and global popularity trends, using bidirectional jump ODEs and variational inference to extract the patterns of information diffusion. In addition, a knowledge distillation loss is incorporated to enhance the consistency of evolution between prior and posterior latent variables. Extensive experiments on real-world datasets demonstrate that VNOIP delivers highly competitive prediction accuracy compared to baselines.

\section{Acknowledgments}
This research was supported by the National Natural Science Foundation of China (Nos.~62261136549, 62471403, U22A2098, 62271411), the Technological Innovation Team of Shaanxi Province (No.~2025RS-CXTD-009), the International Cooperation Project of Shaanxi Province (No.~2025GH-YBXM-017), the Fundamental Research Funds for the Central Universities~(Nos.~G2024WD0151, D5000240309).

\bibliography{aaai2026}

\end{document}